\documentstyle[preprint,tighten,aps,psfig]{revtex}
\begin{document}
\title{Chiral aspects of baryon structure in the quark model.}
\author{ L. Ya. Glozman}
\address{ Institute for Theoretical Physics, University of Graz,
 Universit\"atsplatz 5,\\ A-8010 Graz, Austria; 
e-mail: lyg@physik.kfunigraz.ac.at}
\maketitle

\begin{abstract} 
The implications of chiral symmetry for the quark model
are discussed. In particular its connection with the
meson-baryon approach is outlined. It is stressed that
in the closure approximation, where the intermediate
baryon states are used instead of the original quark basis,
it is imperative to take into account the whole
infinite tower of the radially excited states in order that
the crucially important effects related to the short-range 
part of the meson-exchange interaction be preserved. It is shown 
that the chiral constituent quark model is able to explain the
nucleon axial vector coupling constant and automatically
incorporates all the necessary loop corrections. 
\end{abstract}

\bigskip
\bigskip

It has been suggested some time ago \cite{GR} that the most 
important interactions in 
light and strange baryons in the low-energy regime (below
the chiral symmetry breaking scale $\Lambda_\chi \sim 1$GeV)
are the effective confining interaction and the
short-range part of the Goldstone boson exchange (GBE) interaction
between the constituent quarks, which are quasiparticles
in Bogoliubov or Landau sense and 
related to dynamical chiral symmetry 
breaking. That this 
should be so follows from the fact that the typical momentum 
of valence current quarks in the nucleon is well below the chiral 
symmetry breaking scale, implying that the low-energy characteristics
of baryons, such as their masses, should be formed by the 
nonperturbative QCD dynamics, that is responsible for chiral symmetry
breaking and confinement, but not by the perturbative QCD
degrees of freedom, which become active at much higher momentum scale.\\

From the microscopical side the effective meson exchange
interaction between valence quarks in baryons necessarily
arises from the t-channel iterations of the QCD gluodynamics
which triggers the breaking of chiral symmetry \cite{GV}.
To understand the structure of the nucleon in the low-energy regime it 
is therefore much simpler to use an effective theory, which 
operates with the quasiparticles (in this case constituent quarks and 
chiral fields), which are necessarily implied by the original 
nonperturbative QCD dynamics.\\

The short-range part of the GBE interaction  can be schematically
written as \footnote{
Through the whole paper we use 
pion-exchange. The transition to the whole GBE exchange
within the $SU(3)_F$ limit implies a substitution of the
$SU(2)$ flavor matrices by the $SU(3)$ ones.}

\begin{equation}
-\vec{ \tau}_{i} 
\cdot \vec{\tau}_{j}
\vec{\sigma}_i \cdot \vec{\sigma}_j,
\label{GBE}
\end{equation}

\noindent
multiplied by a radial function. In the extreme case
where the scale of the chiral symmetry breaking approaches
infinity, the radial form of this interaction is given
by the delta-function. In the case of the finite physical
pion and constituent quark sizes it should be some finite function
with the range $\Lambda_\chi^{-1}$.
A short-range force of the same type is also supplied
by the vector-meson-like exchange interaction \cite{GLOZ},
which can also be considered as a representation of the 
correlated two-Goldstone boson exchange, since the latter
has a $\rho$-meson pole in the t-channel \cite{RB}. The same
force is also implied by the Yukawa part of the axial vector
meson exchange. There are phenomenological reasons
to believe that  these contributions also can be  important.
The recent lattice results on the origin of the
$N-\Delta$ splitting \cite{LIU}, large $N_c$ \cite{CARONE}
and phenomenological \cite{GEORGI} analyses of the negative
parity $L=1$ states support the physical picture based on the
interaction (\ref{GBE}).\\

In a recent paper \cite{THOMAS2} Thomas and Krein have
continued a critical discussion of the chiral constituent
quark model. In their first paper \cite{THOMAS} it was argued
that the pion exchange diagram of Fig. 1b, which is the basis
for the interaction (\ref{GBE}), should give results in
conflict with the prediction of baryon chiral perturbation
theory (BChPT). The latter suggests that in the 
$\epsilon$-neighbourhood of the chiral limit the leading
nonanalytic contribution (LNA) to the nucleon mass (i.e. a
contribution of the order $m_c^{3/2} \sim \mu^3$, where $m_c$ 
and $\mu$ are 
current quark and pion masses, respectively) is given by

\begin{equation}
\delta m_N = -\frac{3}{32\pi f_\pi^2}g_A^2 \mu^3 \simeq -15 MeV,
\label{LNA}
\end{equation}

\noindent
where $\mu$ is the pion mass in the neighborhood of
the chiral limit and $g_A$ is
the nucleon axial vector coupling constant in the same regime.
This contribution arises from the nucleon self-energy graph of
Fig. 2a, whereas the diagram of Fig. 2c does not contribute to 
LNA {\it in the chiral limit} \cite{Manohar}. The 
argument made in ref. \cite{THOMAS}
was that the Yukawa part of the pion-exchange interaction of
Fig. 1b,

\begin{equation}
\sim  +\vec{ \tau}_{i} 
\cdot \vec{\tau}_{j}
\vec{\sigma}_i \cdot \vec{\sigma}_j \mu^2 \frac{e^{-\mu r}}{r},
\label{YUKAWA}
\end{equation}

\noindent
would necessarily lead to a splitting of $N$ and $\Delta$
at the order $\mu^3$, which is equivalent to some contribution
of a diagram  of Fig. 1c to the nucleon self-energy at the same 
order. Since
the Yukawa part (\ref{YUKAWA}) of the meson exchange interaction
is firmly related to the short-range part 
(\ref{GBE}) of the pion-exchange interaction, the
authors of ref. \cite{THOMAS} concluded that the splittings
that arise from
this interaction are in conflict with BChPT and
hence wrong.\\

It was, however, proven in ref. \cite{GL}, that the argument
above is incorrect. The reason is that the short-range interaction
(\ref{GBE}) appears at the order $m_c^0 = 1$, i.e. persists in the chiral
limit, while the contribution of the Yukawa part of the meson-exchange
interaction appears only in subleading orders $m_c, m_c^{3/2},...$ and
vanishes in the chiral limit $m_c =0$. Hence, while in the chiral limit
the interaction  (\ref{GBE}) does contribute to the nucleon self-energy
and to the $N-\Delta$ splitting,  the Yukawa part (\ref{YUKAWA}) does not!
Therefore there is no contradiction with  BChPT. Beyond
the chiral limit (i.e. for any finite pion mass $\mu$)
the interaction  (\ref{YUKAWA})  provides
a very small splitting between $N$ and $\Delta$. But in this regime
BChPT states that there also appears a contribution from the diagrams
of Fig. 2c and Fig. 2d  \cite{Manohar,MLAST} and therefore there is no 
contradiction again.\\

In their second paper \cite{THOMAS2} Thomas and Krein suggested
another argument, namely that the baryon-self energy should be obtained
only from the projection onto the intermediate baryon states and that
the $N$ - $\Delta$ splitting, calculated after performing such a
projection, was very different 
to that one implied by Fig. 1a and
Fig. 1b and by the operator (\ref{GBE}).  
Below we show that this new argument
does not hold. In particular we demonstrate that the infinite tower
of the radially excited states must be taken in the meson-baryon
basis in order to incorporate properly the crucially important
effects implied by the interaction (\ref{GBE}). We also show that
the chiral constituent quark model successfully explains the
nucleon axial vector coupling constant and thus correctly incorporates
all the necessary loop corrections.
\\

Consider a generic Hamiltonian of a chiral constituent quark
model, which has also been used by Thomas and Krein:

\begin{equation}
H = H_0 + H_\pi + W,
\label{HAM}
\end{equation}

\noindent
and assume that this Hamiltonian can be treated perturbatively
with respect to the pion-quark interaction $W$ (in reality 
this will not be valid). In the expression above, $H_0$
describes the effective confinement part of the model with
eigenstates $|B^{(0)}_\alpha>$ (called bare quark states
in ref. \cite{THOMAS2}), $H_\pi$ is the Hamiltonian for the noninteracting
pion field. The surprising statement of ref. \cite{THOMAS2} is that
this Hamiltonian could not be treated directly (that would 
indeed correspond
to the sum of diagrams of Fig. 1a and Fig. 1b) but rather an
insertion of a sum over intermediate baryon-pion states should
be done leading to the following self-energy:

\begin{equation}
\Sigma(E) = \sum_n <B_0| W^+ | n> \frac {1}{E-E_n} < n | W | B_0 >.
\label{CL}
\end{equation}

\noindent
In this expression the closure representation for the intermediate
pion-baryon system has been used. It is obvious, however, that the representation
(\ref{CL}) is correct and equivalent to the initial Hamiltonian
(\ref{HAM}) only when the full (infinite) set of the intermediate
 states is taken into account. When the authors
of ref. \cite{THOMAS2} apply this expression to $N$ and $\Delta$
self-energies and obtain the following relations

\begin{equation}
M_N = M_0' - \frac{25}{2} P_{00}' - 16 P_{N\Delta},
\label{MNT}
\end{equation} 

\begin{equation}
M_\Delta = M_0' - 4 P_{\Delta N} - \frac{25}{2} P_{00}', 
\label{MDT}
\end{equation} 

\noindent
they truncate the sum over $n$ in eq. (\ref{CL}) to the ground states
of $N$ and $\Delta$ only. After that they compare the expressions
(\ref{MNT}) and (\ref{MDT}) with those  of \cite{GR} 
(based on the interaction (\ref{GBE}))

\begin{equation}
M_N = M_0  - 15 P_{00}^\pi,
\label{MN}
\end{equation} 

\begin{equation}
M_\Delta = M_0 - 3P_{00}^\pi, 
\label{MD}
\end{equation} 

\noindent
and conclude from the apparent difference that (\ref{MN}) and
(\ref{MD}) should be incorrect. An obvious error is that they
identify $ M_0'$ and  $P_{00}'$ with  $ M_0$ and  $P_{00}^\pi$.
The parameter $M_0$, however, includes
both the contribution from the constituent quark mass and the confinement.
The constituent quark mass, by definition,  contains the
pion self-energy contribution of Fig. 1a \cite{GL}. 
The latter contribution is ultraviolet divergent and is absorbed
into  constituent quark mass after
renormalization.
While
the coefficient $P_{00}'$ takes into account the intermediate
state $N$  in (\ref{MNT}) and  $\Delta$ in (\ref{MDT}),
the coefficient $P_{00}^\pi$ implicitly includes all allowed radially
excited states with all possible radial and 
spin-isospin quantum numbers, which
can be coupled to the nucleon or delta bare ground states via the pion 
absorption (emission) operator. This difference is of crucial
importance.\\

The pion-exchange interaction of Fig. 1b contains both
the ultraviolet part (\ref{GBE}), which is independent of the
pion mass, and the infrared part (\ref{YUKAWA}).  If one is
interested only in the {\it small} effects related with the
infrared part of the meson exchange interaction, as well as in
the infrared contributions from the constituent quark self energy
of Fig. 1 at the order $\mu^3$, then the truncation of the 
intermediate states in
(\ref{CL}) to $N$ and $\Delta$ ground states would be 
approximately justified,
because all other intermediate states will not contribute
at this order.\footnote{ Note again, that this is true only
in the vicinity of the chiral limit. With the real value of the pion mass,
there will also appear some contributions from  radially excited 
states.} 
 However, this is not true for the ultraviolet
part (\ref{GBE}) of the meson exchange interaction, which 
produces a 
much more important effect for the baryon mass.
In this case the truncation of the intermediate states to the ground
$N$ and $\Delta$ states leads to astray. In a perturbative treatment
this is obvious from the fact that the contribution of the interaction 
(\ref{GBE}) is determined by the coefficients $P_{00}^\pi$ which are 
proportional to the square of the bare baryon wave function at {\it zero}
separation between two constituent quarks. If one uses the closure 
representation, one needs a complete infinite set of all possible
radially excited states with all possible spin-isospin quantum numbers
that are allowed by the pion transition in order to represent
correctly the contribution above. Mathematically this 
statement is equivalent to the need for an infinite complete
basis to represent the delta function in the closure representation and  
any effect of the higher terms in this expansion is not smaller 
than the effect
of the first terms. It is easy to verify that the result
obtained with the ground states only, like  in (\ref{MNT})-(\ref{MDT}),
is dramatically different from the correct one in (\ref{MN})-(\ref{MD}).\\

This is also valid for the nonperturbative treatment \cite{GPVW}, 
where the finite
function with the range $\Lambda_\chi^{-1}$ must be used
instead of the original delta function in (\ref{GBE}). For example, 
if one uses a harmonic oscillator
basis in a variational calculation one needs all states up to 
$20\hbar \omega$ to achieve a convergence. This is 
nothing else than taking into account all highly excited intermediate
baryon states.\\ 

Because the identification of $P_{00}'$,  $P_{N\Delta}^\pi$ and
$P_{\Delta N}^\pi$ in (\ref{MNT}) and (\ref{MDT}) with 
$P_{00}^\pi \sim 30$ MeV in (\ref{MN}) and (\ref{MD}) is erroneous,
and the expressions (\ref{MNT}) and (\ref{MDT}) by themselves are
by no means justified, the conclusion about the nucleon
self-energy of $-\frac{57}{2} P_{00}' \sim 855$ MeV is also 
not justified. The correct nucleon self-energy  {\it in the chiral limit} 
is given by $-15P_{00}$ by the expression (\ref{MN}) (which is
due to the interaction (\ref{GBE})). The constituent quark 
self-energy is ultraviolet divergent and thus  completely
absorbed into the definition of the constituent quark mass after
renormalization. The different ultraviolet behaviour of graphs
of Fig. 1a and Fig. 1b is lost in the expressions (\ref{MNT})
and (\ref{MDT}) which do not distinguish between these two graphs.
\\

One may ask, why the typical calculations along  BChPT
ignore all the  intermediate
excited states? It is because the task of BChPT is different. It
considers only corrections that are implied by the nonzero mass
of current quarks (pions) and all contributions  of the chiral
limit (which are numerically much more important) are encoded into
the fitting parameters, which are the tree level baryon 
$N$ and $\Delta$ masses, coupling constants and counterterms.
BChPT  a-priori cannot answer the question  what is the 
origin of the $N-\Delta$ splitting. The task of the quark model
is wider. It should incorporate not only small
corrections from the finite pion mass, but what is much
more important it should offer an explanation of the main contributions
in the chiral limit.\\

When one uses the quark basis and the diagrams of Fig. 1 both the
chiral limit and small corrections from  pion mass are met
properly.
There are obvious advantages even for the treatment
of the small corrections from
the pion mass in this case.
This is because it takes into account explicitly the very different
infrared behaviour of the graphs of Fig. 1a  and Fig. 1b. 
 For example, with the quark basis it is
manifest that the chiral log corrections ($\sim \ln \mu^2$) to the baryon
mass can appear from the quark self-energy graph of Fig 1a only and
they are not supplied by the meson exchange interaction of Fig. 1b
\cite{GL}. This property  is lost in any usual pion-baryon
description (because it does not distinguish between the two graphs
in Fig. 1). It guarantees that when one uses a pion-baryon basis there 
should be
implicit strong cancellations when one considers excited intermediate
states in addition to the ground states, which are ignored in 
the present state of 
BChPT.\\

 In addition,
in the large $N_c$ limit the $SU(6)_{FS}$ symmetry of baryons
becomes exact \cite{DM} (note, that this is not the case for mesons).
This symmetry is also the  symmetry of the nonperturbed basis in the 
quark model for baryons. It is  
crucially important to keep manifest the large $N_c$
behaviour and the $1/N_c$ corrections within the BChPT expansion
because there must be large $N_c$ cancellations of one-loop
corrections, the property which is lost in the usual formulation
of BChPT.
 "Not including the $1/N_c$ cancellations in a systematic
way gives a misleading picture of the baryon chiral expansion" \cite{MLAST}. 
These cancellations are explicit when one uses the quark basis.
The pion exchange interaction of Fig. 1b contributes to the
$N$ and $\Delta$ masses at the order $N_c$, but its contribution to the
$N-\Delta$ splitting and to the violation of $SU(6)$ appears at the
order $1/N_c$, as it must be. 
This implies that the combination of the $SU(6)$ quark basis and
the pionic contributions at the quark level correctly preserves
at the same time the implications of ChPT and large $N_c$ expansion 
at the orders $N_c, 1, 1/N_c$.\\

The next issue to be addressed  is the questioning by the authors
of ref. \cite{THOMAS2} about the statement in ref. \cite{GL} that
the LNA should be mostly due to the constituent quark self-energy
of Fig. 1a. This self-energy implies the following contribution

\begin{equation}
\delta m_N = 3 \times\left\{-\frac{3}{32\pi f_\pi^2}(g_A^q)^2 \mu^3\right\}, 
\label{LNAQ}
\end{equation}

\noindent
which can be easily obtained if one considers the constituent
quarks to be quasifree.\footnote {In reality the constituent quark
is confined and the free Green function should be substituted by
the Green function of a particle in the confining potential. That 
will imply some small modifications of (\ref{LNAQ}).} 
Here $g_A^q$ stands for the axial vector coupling of the constituent
quark.
Using then
the well-known $SU(6)$ relation

\begin{equation}
g_A = 5/3 g_A^q, 
\label{GA}
\end{equation}

\noindent
one obtains a result which is very close to (\ref{LNA}). The 
success of the naive quark model relation (\ref{GA}) suggests that
there should be a deep root in it. 
The $SU(6)$ symmetry of the naive quark model is the exact symmetry
of QCD for baryons in the large $N_c$ limit and in the same limit
the nucleon axial charge is given by the expectation value
of the operator

\begin{equation}
\sum_{l=1}^{N_c} q_l^\dagger \left \{ {\vec \sigma} \otimes {\vec \tau}
\right \} q_l,
\label{G}
\end{equation}

\noindent
evaluated with exact $SU(6)$ wave functions,
which gives a contribution at the order $N_c$ \cite{DM}. 
This procedure, upon
substitution $N_c =3$, results in relation (\ref{GA}).\\

While the relation
(\ref{GA}), which is model independent,  should be considered as
a definition of the constituent quark axial charge in the large
$N_c$ limit {\it with all relativistic and other possible
effects automatically incorporated}, 
it is interesting to learn whether our
intuitive view of constituent quark is compatible with this relation
 and the empirical value for the nucleon axial vector charge, $g_A = 1.25$.\\

The axial vector current of the constituent quark is given as

\begin{equation}
A_\mu^a = ig_A^q {\bar u}(p')\gamma_\mu \gamma_5 \tau^a u(p).
\label{AVC}
\end{equation}

\noindent
In terms of Pauli spinors it reduces to \cite{DANNBOM}

\begin{equation}
A^{ia} = -g_A^q \sigma^i \tau^a \left \{1 - \frac{2}{3}\left (1
-\frac{1}{\sqrt {1+\tilde{{\bf v}}^2}}\right) \right \}.
\label{AVCF}
\end{equation}

\noindent
Here the "velocity" operator  is defined as 
$\tilde{{\bf v}} = ({\bf p}' + {\bf p})/2m$. This "velocity"
operator should not be mixed with the standard one-particle
relativistic velocity ${\bf v}$. In the limit ${\bf p}' = {\bf p}$
they are obviously related via 
${\tilde {v}}^2 = \frac{v^2}{1-v^2}$.
Note that the expression (\ref{AVCF})  is
exact and does not assume a nonrelativistic expansion of the
constituent quark spinors. It is explicitly seen from this
expression that the relativistic motion factor in  curly brackets
{\it can be absorbed into a definition of the 
 axial vector charge of the constituent quark that is effectively at rest}
and thus one indeed obtains the recipe (\ref{G}).
Assuming that the constituent quark is  relativistic, 
$v^2 \simeq  0.5 \div 0.7$
(as it indeed
follows from  the semirelativistic dynamical calculations \cite{GPVW} 
with the confining interaction  as the only interaction between
constituent quarks),  one obtains
relativistic motion correction factor $0.8 \div 0.7$. 
This factor, combined
with expression (\ref{GA}) gives 
$g_A = 5/3 \times 0.8 \div 0.7 = 1.3 \div 1.2$. In this
calculation it has been assumed that 
in the large $N_c$ limit the "bare" constituent
quark axial vector charge $g_A^q$ in  (\ref{AVCF})
is given by 1 \cite{WEIN}
(there are some arguments, however, that it should not  
necessarily be so
\cite{PERRIS}).\\
 
 All  correction to the expression (\ref{GA}) start 
at the order $1/N_c$ \cite{DM}. To these belong corrections
due to the pion cloud of constituent quarks \cite{WEIN2}
and due to the pion exchange interaction.
 As explained in ref. \cite{GL}
and above the pion exchange interaction, which contributes
to the violation of $SU(6)$ at the order $1/N_c$, results in the
configuration mixing at this order and provides thus a small
renormalization of the expression (\ref{GA}). It also  contains the 
$\sim \mu^3$ contribution to the nucleon mass 
 and thus 
provides other small renormalization of the expression (\ref{LNAQ}).
The latter contributions represent effects of the quark enviroment
because they are  two-body contributions and are determined by the
quantum numbers of a pair of quarks, but not of a single isolated 
quark. 
Since the model does inherently contain the loop contributions of Fig. 2 
and is able to reproduce (at least approximately) the axial charge
of the nucleon, it necessarily reproduces  LNA.\\ 

Finally a comment on the remarks by Thomas and Krein \cite{THOMAS2} 
and Lipkin
\cite{LIPKIN}  regarding the failure of the prediction of the
naive constituent quark model for some specific combination
of $\Sigma$ and $\Lambda$ magnetic moments. It is not surprising
that the naive constituent quark model, which ignores all effects 
implied by  chiral symmetry dynamical breaking 
fails here. Within the chiral constituent quark model these implications
assume the pion-loop contribution at the quark level \cite{GR2} 
as well as the meson exchange current
(MEC) contributions (see e.g. the recent papers \cite{DANNBOM,MEC}). 
The latter represent  two-body 
effects, i.e. effects of the quark enviroment (in the language of 
Thomas and
Krein \cite{THOMAS2} and papers cited therein). While the results of
the qualitative calculations  show that the
effects of the MEC for some observables,  specifically for the 
magnetic moments should be significant  and work in a proper
direction \cite{DANNBOM}
(because, as it is well known from nuclear physics, the MEC 
contributions to M1 electromagnetic transitions are large), one cannot
unfortunately calculate them reliably because the problem is
highly relativistic and the derivation of the meson exchange
current operators at the order  $v^2/c^2$, justified in nuclear
physics, cannot be considered to be quantitative in the present
application. But an important issue, which we want to stress,
is that the LNA contribution and other chiral corrections
to magnetic moments {\it are taken into account} as soon as
pion loops and MEC contributions are properly included.
\\

The conclusion is that the meson-baryon or quark models
which employ only the subspace of the pion-nucleon and 
pion-delta states, 
are incomplete and ignore the most
important short range effects of the pion (meson) degrees of 
freedom for the baryon
structure. If one uses the pion(meson)-baryon basis it is imperative to 
include the whole tower of the excited baryon states.\\

The author is indebted to D.O. Riska and W. Plessas for 
carefully reading  the manuscript.

{\bf Figure captions}

Fig.1 Pion loop a) and pion exchange b) contributions to the
baryon mass within the chiral constituent quark model.

\bigskip

Fig.2 Pion loop contributions to the baryon mass with the
pion-baryon system in the intermediate state truncated to $N$ and $\Delta$
states only.

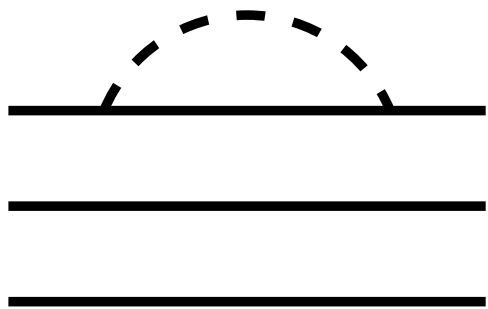
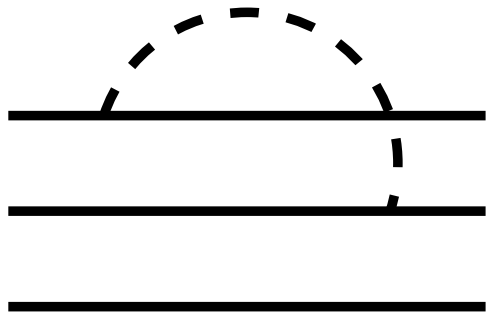
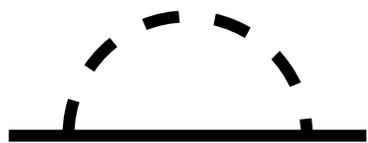
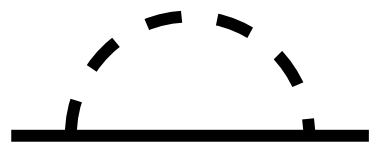
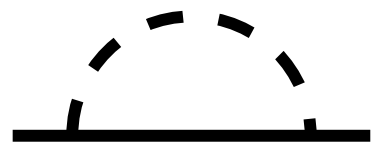
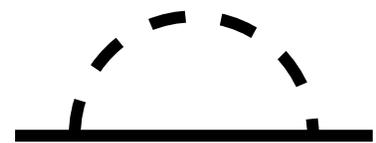
\end{document}